\documentclass[12pt,a4paper]{article}
\pdfoutput=1
\usepackage{graphicx}
\usepackage{xspace}
\usepackage{jheppub}
\usepackage{dsfont}
\usepackage{hyperref}
\usepackage{relsize,exscale,scalefnt,anyfontsize,cases}
\usepackage[utf8]{inputenc}

\usepackage{bm}
\usepackage{multirow}

\newcommand{\be}{\begin{equation}}
\newcommand{\ee}{\end{equation}}
\newcommand{\bea}{\begin{eqnarray}}
\newcommand{\eea}{\end{eqnarray}}
\newcommand{\dd}{\textnormal{d}}
\newcommand{\rpa}{\ensuremath{R_{\rm pA}}\xspace}
\newcommand{\xf}{\ensuremath{x_{\textnormal{F}}}\xspace}
\newcommand{\qzero}{\ensuremath{\hat{q}_0}\xspace}
\newcommand{\gevsqfm}{\ensuremath{\textnormal{GeV}^2/\textnormal{fm}}\xspace}
\newcommand{\gev}{\ifmmode \xspace\ensuremath{\textnormal{GeV}\xspace} \else GeV \fi}
\newcommand{\qhat}{\hat{q}}
\def\xone{\ensuremath{x_{1}}\xspace}
\def\xtwo{\ensuremath{x_2}\xspace}
\newcommand{\ie}{{i.e.}\xspace}
\newcommand{\eg}{{e.g.}\xspace}
\newcommand{\alphas}{\alpha_s}
\newcommand{\A}{{\rm A}}
\def\cO#1{{{\cal{O}}}\left(#1\right)}

\newcommand{\jpsi}{\ensuremath{{J}\hspace{-.08em}/\hspace{-.14em}\psi}\xspace}
\newcommand{\eq}[1]{Eq.~\eqref{#1}\xspace}
\newcommand{\pp}{{\ensuremath{\textnormal{pp}}}\xspace}
\newcommand{\pA}{\ensuremath{\textnormal{pA}}\xspace}
\newcommand{\hp}{\ensuremath{\textnormal{hp}}\xspace}
\newcommand{\hA}{\ensuremath{\textnormal{hA}}\xspace}

\newcommand{\piA}{\ensuremath{\pi\textnormal{A}}\xspace}
\newcommand{\sqrts}{\ensuremath{\sqrt{s}}\xspace}
\newcommand{\pt}{\ensuremath{p_{_\perp}}\xspace}
\newcommand{\dptsq}{\ensuremath{\Delta p_{_\perp}^2}\xspace}
\newcommand{\dptsqfcel}{\ensuremath{\dptsq\big|_{\rm FCEL}}\xspace}
\newcommand{\dptsqnpdf}{\ensuremath{\dptsq\big|_{\rm nPDF}}\xspace}
\newcommand{\ptsq}{\ensuremath{\langle p_{_\perp}^2} \rangle\xspace}
\newcommand{\meanptsq}{\ensuremath{{\langle \pt^2 \rangle}}\xspace}
\newcommand{\lcoh}{\ensuremath{\ell_{\textnormal{coh}}}\xspace}
\newcommand{\xip}{\ensuremath{{x}_{\textnormal{p}}}\xspace}
\newcommand{\xiA}{\ensuremath{{x}_{\A}}\xspace}
\newcommand{\chindf}{\ensuremath{\chi^2/\textnormal{ndf}}\xspace}

\title{Nuclear {\boldmath $p_\perp$}-broadening of Drell-Yan and quarkonium production from SPS to LHC}

 \author[a]{Fran\c{c}ois Arleo,}
 \author[a,b]{Charles-Joseph Na\"{i}m}

\affiliation[a]{Laboratoire Leprince-Ringuet, \'Ecole polytechnique, Institut polytechnique de Paris,\\ CNRS/IN2P3, 91128 Palaiseau, France}
\affiliation[b]{IRFU, CEA, Universit\'e Paris-Saclay, 91191 Gif-sur-Yvette, France}
 \emailAdd{francois.arleo@cern.ch}
 \emailAdd{charles-joseph.naim@cern.ch}

\abstract{The nuclear $p_\perp$-broadening of Drell-Yan and quarkonium ($J/\psi$, $\Upsilon$) production in $\pi$A and pA collisions is investigated. The world data follow a simple scaling from SPS to LHC energies, once the process-dependent color factors are properly taken into account, which allows for the extraction of the transport coefficient in cold nuclear matter. We find that $\qhat(x)\propto x^{-\alpha}$ with $\alpha=0.25$--$0.30$. The magnitude of the transport coefficient at $x=10^{-2}$ is $\qzero=0.051$~\gevsqfm and $\qzero=0.075$~\gevsqfm, whether $Q\bar{Q}$ pairs are assumed to be produced as color octet or color singlet states, respectively. The relation between nuclear broadening data and the (CT14) gluon density is also investigated.}

\begin{document} 
\maketitle
\flushbottom

\renewcommand*{\thefootnote}{\arabic{footnote}}
\setcounter{footnote}{0}

\section{Introduction}\label{sec:Introduction}

The multiple scattering incurred by energetic quarks and gluons propagating in a QCD medium is responsible for both induced gluon radiation --~leading to parton energy loss~-- and transverse momentum broadening. While the former is responsible for the jet quenching phenomena discovered in heavy ion collisions at RHIC and at LHC (see Refs.~\cite{Majumder:2010qh,Mehtar-Tani:2013pia,Armesto:2015ioy,Qin:2015srf} for reviews), observing the latter is more delicate. Measuring dijet azimuthal correlations in proton-nucleus~\cite{Adam:2015xea,Aaboud:2019oop} and nucleus-nucleus collisions could in principle allow for probing transverse momentum broadening, respectively in cold nuclear matter and hot quark-gluon plasma~\cite{Kharzeev:2004bw,Dominguez:2011wm,Mueller:2016gko,vanHameren:2019ysa}. In practice, however, these correlations may prove more sensitive to medium-independent Sudakov radiation rather than to multiple scattering in a medium and thus deserve detailed scrutiny~\cite{Mueller:2016xoc}.

A perhaps simpler observable, namely the nuclear broadening of \emph{single inclusive} particle production in \hA collisions,
\be\label{eq:broadeningdef}
\dptsq = \meanptsq_{\hA} - \meanptsq_{\hp}\,,
\ee
offers a direct access to the saturation scale, hence a measure of multiple scattering in nuclei~\cite{Baier:1996sk,Kopeliovich:2010aa,Duraes:2015qoa}.
In this article, we perform a systematic study of Drell-Yan (DY) and quarkonium ($\jpsi$, $\Upsilon$) nuclear broadening data in \piA and \pA collisions, from fixed-target (SPS, FNAL) to collider (RHIC, LHC) energies. It is found that all measurements, once scaled with appropriate color factors, exhibit a simple dependence expected from perturbative QCD. This data-driven analysis allows in turn for the determination of the transport coefficient in cold nuclear matter, or equivalently the saturation scale in large nuclei and its energy dependence.

The model for the nuclear broadening is presented in section~\ref{sec:model} before results are shown and discussed in section~\ref{sec:results}. Conclusions are drawn in section~\ref{sec:conclusion}.

\section{Model and data}\label{sec:model}
\subsection{Transverse momentum broadening}\label{sec:broadening}
\subsubsection*{Asymptotic parton}
The accumulated transverse momentum acquired by an asymptotic parton (in color representation $R$) traversing a medium of length $L$ is given by~\cite{Baier:1996sk}
\be
\dptsq = \qhat_R\,L\,,
\ee
where $\qhat_R$ is the transport coefficient of that medium.\footnote{Logarithmic corrections, $\sim \ln L$ and $\sim \ln^2 L$, to the broadening or to the transport coefficient~\cite{Iancu:2014kga,Blaizot:2014bha} are here ignored.}  It is given by $\mu^2 / \lambda_R$ where $\mu$ is the typical transverse momentum exchange between the parton and each scattering center, and $\lambda_R$ is the parton mean free path in the medium. Defining $\qhat$ as the \emph{gluon} transport coefficient and using $C_R\,\lambda_R\ = N_c\,\lambda_g$, the broadening becomes
\be\label{eq:broadening}
\dptsq = \frac{C_R}{N_c}\,\qhat\,L\,,
\ee
where $C_R$ is the color charge of parton $R$ ($C_F=(N_c^2-1)/2N_c$ for a quark, $C_A=N_c$ for a gluon).

\subsubsection*{Hard QCD process on a nuclear target}

\begin{figure}[h!]
    \centering
    \includegraphics[width=9.cm]{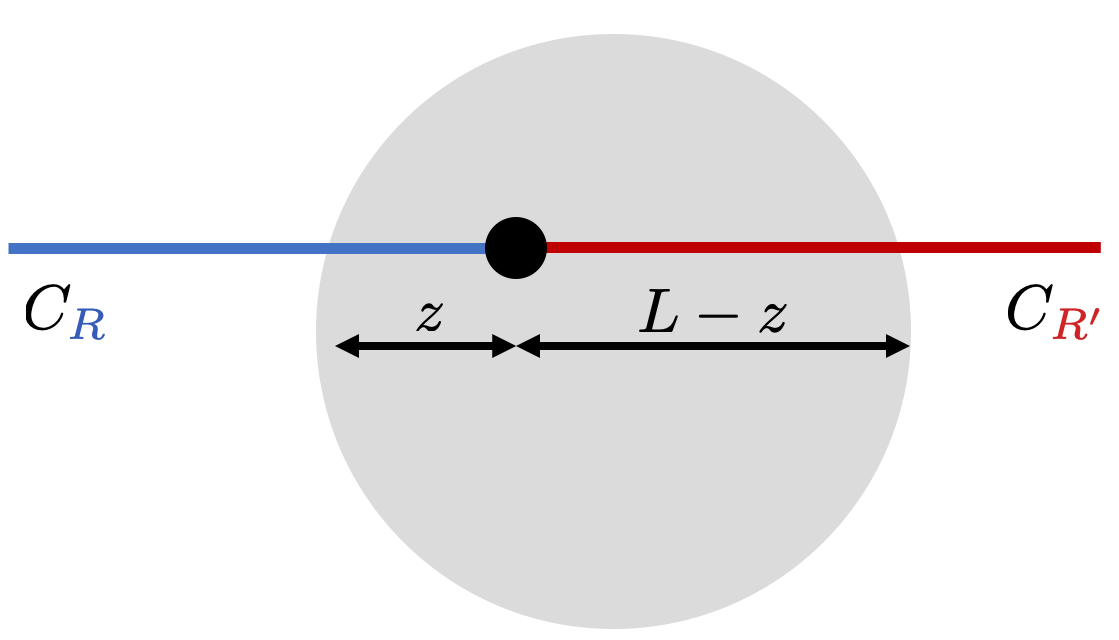}
    \caption{Sketch of the nuclear broadening at small coherence length.}
    \label{fig:sketch}
\end{figure}

Let us now consider the more realistic case of the production of a single massive particle produced by a hard QCD process in a hadron-nucleus collision.\footnote{More explicitely, we will only address the production of Drell-Yan massive lepton pairs and quarkonia ($\jpsi$ and $\Upsilon$), in \pA and \piA collisions.}
When the coherence length of the hard process is small, typically smaller than the internucleonic distance, $\lcoh \lesssim 1$~fm, the process as seen in the nuclear rest frame can be sketched as in Figure~\ref{fig:sketch}: (i) an incoming parton (in color representation $R$) stemming from the incoming hadron experiences transverse momentum broadening over a distance $z$, at which it participates to the hard process; (ii) an outgoing particle (in color representation $R^\prime$) is produced on a short length scale $\lcoh$ and propagates in the medium over a length $L-z$, until it exits the nucleus. 

In this picture, the nuclear transverse momentum broadening in \hA collisions compared to \hp collisions is simply given by\footnote{As we shall see in section~\ref{sec:qhat}, the transport coefficient acquires a mild dependence on the medium length, hence the subscripts $\qhat_\A$ and $\qhat_{\textnormal{p}}$.}
\be\label{eq:broadeningpa}
\dptsq = \frac{\cal C}{N_c}\, \left( \qhat_\A\,L_\A - \qhat_{\textnormal{p}}\,L_{\textnormal{p}} \right)\,,
\ee
where the color factor ${\cal C}$ is given by the half sum of the Casimir factors of the initial and final-state particle,\footnote{As shown in Ref.~\cite{Cougoulic:2017ust}, \eq{eq:colorfactor} holds quite generally for any incoming and outgoing \emph{pointlike} multi-parton state in any representation $R$ and $R^\prime$.}
\be\label{eq:colorfactor}
{\cal C} = \frac{C_R + C_{R^\prime}}2\,.
\ee
Assuming a hard sphere nuclear density of radius $R_\A = 1.12\times A^{1/3}$~fm, the average medium length in a large nucleus is given by $L_\A = 3/2\, R_\A$ after averaging uniformly over the position of the production process. The length scale in a proton target is set to $L_\textnormal{p}=1.5$~fm, as in Refs.~\cite{Arleo:2012hn,Arleo:2012rs}.

In the limit of large coherence length, $\lcoh \gg L$, the broadening given by \eq{eq:broadeningpa} with the generic color rule~\eq{eq:colorfactor} may not hold for all hard QCD processes. In that limit, the outgoing particle (in the present context, a virtual photon or a compact $Q\bar{Q}$ pair) detected experimentally would occur, at the amplitude level, from the radiation of the incoming parton either long before or long after crossing the nuclear target in its rest frame. In the case of initial-state (respectively, final-state) radiation in both the amplitude and complex conjugate amplitude, the broadening would depend solely on the Casimir of the outgoing particle $C_{R^\prime}$ (respectively, that of the initial parton $C_{R}$), while the broadening associated to the interfering terms between initial and final state radiation would have no simple expression in terms of Casimir factors. Consequently, \eq{eq:broadeningpa} may not be appropriate to describe the broadening of Drell-Yan lepton pairs at high energy, as the initial parton (a quark) and the outgoing particle (the virtual photon) carry different color charges. In any case, measurements of DY broadening are not available yet at large coherence length, \ie at RHIC and LHC, despite the preliminary results being reported by the PHENIX experiment~\cite{Leung:2018tql} which are not included in this analysis. On the contrary, when the incoming and outgoing particles carry the same color charge, as would be the case for an octet $Q\bar{Q}$ pair induced by an incoming gluon ($C_R=C_{R^\prime}$), \eq{eq:broadeningpa} is likely to hold (\ie, ${\cal C}=N_c$). We shall therefore assume in the following that the broadening of quarkonia at high energy (RHIC, LHC) is given by \eq{eq:broadeningpa}.

\subsection{Process dependence}\label{sec:processdependence}

In this section we discuss the specific color factor ${\cal C}$ expected for Drell-Yan and quarkonium production in \piA and \pA collisions.

\subsubsection*{Drell-Yan}

At Born level, DY lepton pairs are produced in hadron-nucleus collisions by the annihilation of a quark (respectively, an antiquark) from the projectile hadron with an antiquark (respectively, a quark) from the nuclear target. Since the virtual photon (or the lepton pair) does not experience multiple scattering in the target, $C_{R^\prime}=0$, the expected color factor is thus ${\cal C}_{_{\textnormal{DY}}} = C_F / 2$, independently of the nature of the projectile hadron (here, a pion or a proton). At order $\cO{\alphas}$, the QCD Compton process, $q g \to q \gamma^\star$, may come into play. However, it is not expected to dominate the cross section at low $\pt \lesssim M_{_{\textnormal{DY}}}$, where $M_{_{\textnormal{DY}}}$ is the invariant dilepton mass. Moreover, at forward dilepton rapidity (as is the case for the DY data presently analyzed), this process is dominated by the fusion of an incoming \emph{quark} with a gluon from the target (as $q(\xone) g(\xtwo) \gg g(\xone) q(\xtwo)$) leading to the same color factor as for the $q\bar{q}$ annihilation process.

\subsubsection*{Quarkonium}

Which color factor to use for quarkonium production is more delicate for two reasons. At leading order in $\alphas$, both gluon fusion ($gg\to Q\bar{Q}$) and quark-antiquark annihilation ($q\bar{q}\to Q\bar{Q}$) processes are expected to contribute to heavy-quark pair production, which would lead, respectively, to $C_R=N_c$ and $C_R=C_F$ in \eq{eq:colorfactor}. In \pA collisions, however, quarkonium production should be dominated by gluon fusion unless the longitudinal momentum fraction becomes very large, $\xf\gtrsim0.5$. This statement depends mostly on the momentum distributions of quarks and gluons, and appears to be true in non-relativistic QCD (NRQCD) or in the Color Evaporation Model (CEM)~\cite{Vogt:1999dw}, leading to $C_R=N_c$ in \eqref{eq:colorfactor}. The $q\bar{q}$ annihilation process is more significant in \piA collisions as the pion carries valence \emph{antiquarks}. Using the pion PDF recently extracted in Ref.~\cite{Barry:2018ort} in the CEM at leading order, we find that the $q\bar{q}$ annihilation  channel actually dominates the inclusive $\jpsi$ production in \piA collisions (hence, $C_R=C_F$), at least in the $\xf$ range considered in this analysis.

\begin{table}[t]
  \begin{center}
    \begin{tabular}{p{4.5cm}cc}
      \hline 
      \hline 
      Process & Collision  & ${\cal C}$\\
      \hline
      Drell-Yan & \piA / \pA  & $C_F/2$\\
      Quarkonium & \piA  & $(C_F+N_c)/2$ \\
      Quarkonium  & \pA  & $N_c$\\[0.1cm]
      \hline \\[-0.4cm]
      Quarkonium (singlet)  & \piA / \pA  & $N_c/2$ \\
      \hline 
      \hline
    \end{tabular}
     \caption{Color factors assumed in the present analysis for Drell-Yan and color octet quarkonium production in \piA and \pA collisions. The assumption regarding the case of color singlet quarkonium production is discussed in section~\ref{sec:colorsinglet}.}
    \label{tab:color}
  \end{center}
\end{table}

In addition, the quarkonium production process is still poorly understood. In particular it is not clear on which length scale ($\ell_{\textnormal{octet}}$) the color octet heavy-quark pair neutralizes its color, either $\ell_{\textnormal{octet}} \sim \lcoh$ (as in the Color Singlet Model) or $\ell_{\textnormal{octet}} \gg \lcoh$ (NRQCD, CEM). In this study we shall assume that the $Q\bar{Q}$ pair remains in a color octet state during its entire propagation throughout the nucleus, leading to $C_{R^\prime}=N_c$. For completeness, the assumption of color singlet $Q\bar{Q}$ production is also discussed in section~\ref{sec:colorsinglet}.
The color factors assumed in this study for DY and quarkonium production in hadron-nucleus collisions are summarized in Table~\ref{tab:color}. 

\subsection{Transport coefficient}\label{sec:qhat}

Apart from the Casimir scaling properties discussed in the previous section, the other crucial ingredient which governs the nuclear broadening of hard QCD processes is $\qhat$, the gluon transport coefficient in cold nuclear matter. It is related to the gluon distribution $xG(x, Q^2)$ inside each nucleon of the target~\cite{Baier:1996sk},
\be\label{eq:qhatBDMPS}
\qhat(x, Q^2) = \frac{4 \pi^2 \alphas(Q^2) N_c}{N_c^2 - 1}\,\rho\,xG(x, Q^2)\,,
\ee
where $\rho$ is the nuclear density.

In the case of a hard parton produced inside the medium (consider for instance a quark produced in deep inelastic scattering at large Bjorken-$x$), typically when the coherence length is small, $\lcoh \lesssim L_{\A}$, the value of $x$ at which the gluon distribution needs to be probed is $x \sim \xiA \equiv 1/(2 m_{_{\textnormal{N}}} L_{\A}) > 10^{-2}$~\cite{Baier:1996sk}. On the contrary, when the hard parton is produced by a QCD process coherent over the whole nucleus, $\lcoh \gg L_{\A}$, $x$ should be given by $x \sim 1/(2 m_{_{\textnormal{N}}}\, \lcoh) = \xtwo$, where $\xtwo$ is the momentum fraction carried by the parton probed in the nuclear target~\cite{Arleo:2012rs}. The scale $Q^2$ which controls both the running coupling $\alpha_s$ and the QCD evolution of $xG$ should be of the order of the broadening itself, $Q^2 \sim \dptsq$, yet in practice this semi-hard scale should be frozen when $\dptsq$ becomes too small, $\dptsq \lesssim 1$--$2$~GeV$^2$.

At small $x \ll 10^{-2}$, the gluon distribution exhibits a power law behavior, $xG(x, Q^2) \propto x^{-\alpha}$, where the exponent $\alpha\simeq0.2$--$0.3$ depends only slightly on the resolution scale $Q$. Neglecting the $Q^2$ dependence of $\qhat$ in \eq{eq:qhatBDMPS}, the transport coefficient can thus be modelled as~\cite{Arleo:2012rs}\footnote{In Ref.~\cite{Arleo:2012rs} the exponent $\alpha=0.3$ was assumed.}
\be \label{eq:qhatmodel}
\qhat_\A(x) = \qzero\times\left(\frac{10^{-2}}{x}\right)^{\alpha}
\quad;\quad x=\min(x_\A, \xtwo)\,.
\ee
where $\qzero$ is the transport coefficient at $x = 10^{-2}$. Note that at large $\xtwo > x_\A$, $\qhat$ acquires a mild dependence on the size of the nucleus from the $L$ dependence of $x_A$. 
Casting \eqref{eq:qhatmodel} in \eqref{eq:broadeningpa} leads to
\be\label{eq:qhatmodel2}
\dptsq = \frac{\qzero}{N_c} \times \left(\frac{10^{-2}}{x}\right)^\alpha \times {\cal C}\,\Delta L \quad;\quad \Delta L = L_A - L_p^\prime 
\ee
where $L_p^\prime$ is defined as $L_p^\prime = L_p \left(\frac{\min(x_\A, \xtwo)}{\min(x_{\rm p}, \xtwo)}\right)^\alpha$. At small $\xtwo < x_\A$, $L_p^\prime$ coincides with $L_p$, while at large $\xtwo > x_p$ the factor accounts for the fact that the transport coefficient should be evaluated at two different values of $x$ in the proton and in the nucleus targets. The value of $\xtwo$ is given by $\xtwo=M/\sqrt{s}\,e^{-y}$. In the present approach, the absolute magnitude $\qzero$ and the slope $\alpha$ are the only parameters which need to be extracted from the measurements.
The relationship \eqref{eq:qhatBDMPS} between the nuclear broadening and more realistic gluon distribution functions will be investigated in section~\ref{section:xGx}.

\subsection{Other nuclear effects}\label{sec:fcelnpdf}

It has been assumed so far that parton multiple scattering in nuclei is the only process which leads to finite nuclear $\pt$-broadening, $\dptsq\neq0$. However, other cold nuclear matter effects such as fully coherent energy loss (FCEL)~\cite{Arleo:2010rb,Peigne:2014uha,Peigne:2014rka} or nuclear parton distribution functions (nPDF)~\cite{Armesto:2006ph,Eskola:2009uj,deFlorian:2011fp,Eskola:2016oht,Kovarik:2015cma} may distort the shape of particle $\pt$-spectra in hadron-nucleus collisions with respect to hadron-proton collisions.

The $\pt$ dependence of the nuclear production ratio,
\begin{equation}
    \label{eq:DY_ratio}
    R_{\textnormal{hA}}(\pt) =  \frac{1}{A}\,\frac{\dd\sigma_{\hA}(\pt)}{\dd \pt} \,\Big/\,  \frac{\dd\sigma_{\hp}(\pt)}{\dd \pt},
\end{equation}
due to fully coherent energy loss has been investigated in the case of quarkonium~\cite{Arleo:2013zua} and light hadron~\cite{Arleo:2020eia,Arleo:2020hat} production. The average fully coherent energy loss is suppressed by one power of the hard scale, $\Delta E \propto 1/\pt$ at $\pt\gg M$, making $\rpa(\pt)$ a growing function of $\pt$ in the rapidity intervals presently discussed. Consequently, the quarkonium $\pt$-spectra prove harder in hA collisions than in hp collisions, thus leading to a positive `nuclear broadening', $\dptsqfcel>0$. Unlike quarkonium production, no FCEL is expected in the Drell-Yan channel~\cite{Arleo:2015qiv}, which may nonetheless be sensitive to initial-state energy loss. These effects, however, prove tiny either when $\xf$ is not too large or at high collision energies~\cite{Arleo:2018zjw}.

Let us now discuss the role of nPDF effects on nuclear $\pt$-broadening. At large $\pt \gtrsim M$, the momentum fraction probed in the nuclear target depends on the transverse momentum of the detected particle, $\xtwo = \xtwo(\pt)$. Assuming for simplicity that only one parton species from the nucleus (call it flavour $i$) dominates the cross section, the $\pt$ spectrum in $\hA$ collisions may be given by (neglecting here broadening effects)
\be\label{eq:npdf}
\frac{1}{A} \frac{\dd\sigma_\hA^{\textnormal{nPDF}}(\pt)}{\dd \pt} = R_i^\A\left(\xtwo(\pt), Q(\pt)\right) \times \frac{\dd\sigma_\hp(\pt)}{\dd \pt}\,,
\ee
where $R_i^\A(x, Q) \equiv f_i^\A(x, Q) / A\,f_i^{\textnormal{p}}(x, Q)$ is the nPDF ratio of parton flavour $i$ in the nucleus \A\ over that in a proton, and the factorization scale $Q$ might itself depends on $\pt$. When $R_i^\A$ varies little with \pt compared to the weighted spectrum $\pt^2\,\dd\sigma_\hp/\dd \pt$, the shape of the spectrum is not affected by nPDF corrections but only its magnitude. As a consequence, no net effect of nuclear parton densities on the broadening is expected. This may be the case in the deep shadowing region at very small $\xtwo$, typically $\xtwo \lesssim 10^{-4}$,  and in the vicinity of the anti-shadowing region, $0.05 \lesssim \xtwo \lesssim 0.2$.\footnote{Of course these values are indicative as they may vary from one nPDF set to another.} In between these two domains, $R_i$ is often a \emph{growing} function of $\xtwo$ (and thus of $\pt$), making the particle spectrum harder in \hA collisions compared to \hp collisions. As a consequence, the sole nPDF effects would lead to a positive contribution to the nuclear broadening, ${\dptsq}\big|_{\textnormal{nPDF}} > 0$, if the typical transverse momenta contributing to $\ptsq$ correspond to these domains in $\xtwo$. Conversely, at larger $\xtwo \gtrsim 0.2$, the possible decrease of $R_i^{\A}$ with $\xtwo$ due to the EMC effect would lead to softer spectra in \hA collisions, leading to a \emph{negative} contribution to the nuclear broadening, ${\dptsq}\big|_{\textnormal{nPDF}} < 0$. At fixed-target collision energies, for which $\pt \lesssim M$, the momentum fraction depends only mildly on $\pt$, therefore no strong nPDF effects on the broadening are expected. 

Quantitative results on the FCEL and nPDF effects to the nuclear broadening are discussed in Appendix~\ref{app:fcelnpdf}. These prove significantly less than in data, which confirms that the measured nuclear broadening can be mostly attributed to multiple scattering effects, rather than to FCEL or nPDF.

\subsection{Data}\label{sec:data}

In the present analysis, all available data on the nuclear broadening of Drell-Yan, \jpsi, and $\Upsilon$ production have been used, from SPS energy ($\sqrts \approx 20$~GeV) to the top LHC energy ($\sqrts=8.16$~TeV).
\begin{table}[t!]
    \centering
    \begin{tabular}{cccccc}
      \hline 
      \hline 
      Exp. & Proj.  & Target & $\sqrt{s}$ (GeV) & Process & Ref. \\
      \hline 
      NA3  &  p & Pt & 19.4  & $\jpsi$  & \cite{Badier:1983dg} \\
           & $\pi^{-}$ & Pt & $16.8 / 19.4 / 22.9$   & $\jpsi$& \\
           & $\pi^{+}$ & Pt & 19.4    & $\jpsi$& \\[0.3cm]
      \hline 
      NA10 & $\pi^{-}$ & W & $16.2 / 23.2$  & DY  &   \cite{Bordalo:1987cr} \\ 
           & $\pi^{-}$& W & $23.2$    & $\jpsi$ & \\[0.3cm]
      \hline 
         E772    & p & Ca, Fe, W & 38.7    & DY & \cite{McGaughey:1999mq}\\
           & p  & Ca, Fe, W       & 38.7 &   $\Upsilon$ &\\[0.3cm]
     \hline 
         PHENIX  & d &  Au  & 200 & $\jpsi$ & \cite{Adare:2012qf} \\[0.3cm]
      \hline 
      ALICE & p &  Pb & 5020  &$\jpsi$ & \cite{Adam:2015jsa} \\[0.3cm]
      \hline 
      LHCb & p & Pb & 8160   & $\jpsi$ & \cite{Aaij:2017cqq} \\[0.3cm]
      \hline 
      \hline
    \end{tabular}
     \caption{Data sets included in the present analysis.}
    \label{tab:datasets}
\end{table}

The analyzed data sets are summarized in Table~\ref{tab:datasets}. At SPS (NA3, NA10), measurements have been performed in \piA and \pA collisions, which allows for investigating the color charge dependence of \jpsi nuclear broadening; see Table~\ref{tab:color}. The E772 results performed in \pA collisions on different nuclear targets are sensitive to the path-length dependence of both DY and $\Upsilon$ nuclear broadening. Finally, the RHIC and LHC measurements are carried out on a single nuclear target (Au and Pb, respectively) but in different rapidity ranges, thus probing the \xtwo dependence of the transport coefficient, as emphasized \eg in Ref.~\cite{Duraes:2015qoa}. In particular, the forward measurements at RHIC ($1.2 < y < 2.2$) and at LHC ($2 < y < 4$, for the LHCb experiment) correspond to data taken at the smallest \xtwo values, $\xtwo \approx 3\times10^{-3}$ and $\xtwo \approx 2\times10^{-5}$, respectively. 

At the LHC, the \jpsi nuclear \pt-broadening in minimum bias \pA collisions have not been released by the ALICE and LHCb collaborations. The extraction of $\dptsq$ from the measurements of the absolute $\pt$-spectra, as well as a new extraction from PHENIX data~\cite{Adare:2012qf}, are detailed in Appendix~\ref{app:broadening}.

\section{Results}\label{sec:results}

\subsection{Scaling}\label{sec:scaling}

\begin{figure}[tbp]
    \centering
    \includegraphics[scale=0.55]{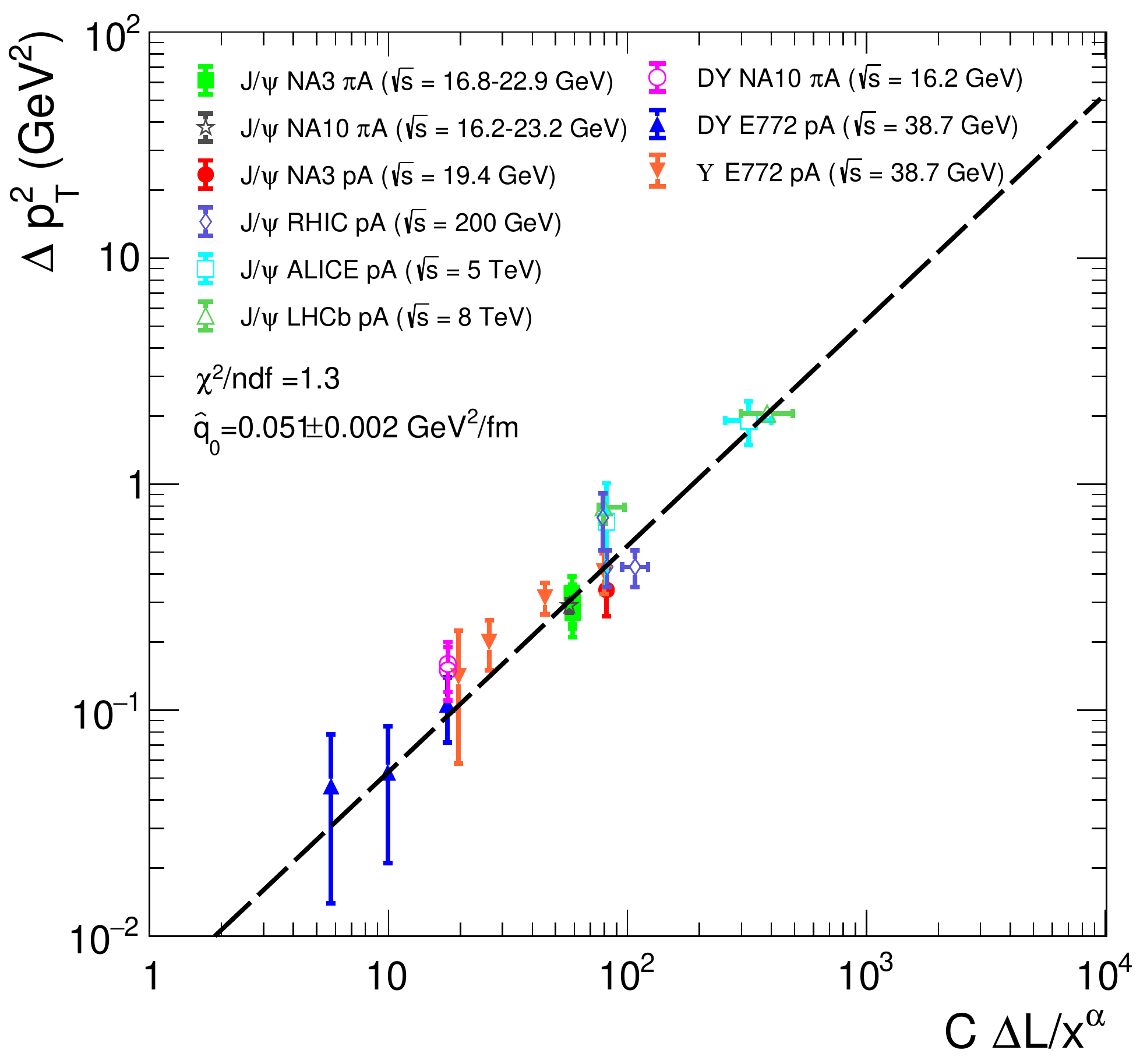}
    \caption{Scaling of the nuclear \pt-broadening in DY and quarkonium production using the color assumptions in Table~\ref{tab:color}.}
    \label{fig:scalingCEM}
\end{figure}

In order to test the above picture, the world data on nuclear transverse momentum broadening of Drell-Yan, $\jpsi$ and $\Upsilon$ production is plotted in Figure \ref{fig:scalingCEM} as a function of ${\cal C}\,\Delta L / x^\alpha$, see~\eq{eq:qhatmodel2}. 
Remarkably, the measurements performed in \pA and \piA collisions, for different particle species and on a large range of collision energies from SPS to LHC, exhibit clearly the scaling property. As expected, the lowest values of $\dptsq$ are obtained in the DY process (${\cal C}=C_F/2$) at low collision energy (hence, larger $\xtwo$ for which $\qhat$ is lower). On the contrary, the largest broadening is observed for quarkonium production at the LHC, because of both the large color factor (${\cal C}=N_c$) and the small-$\xtwo$ rise of the gluon distribution, $\qhat(\xtwo) \propto \xtwo^{-\alpha}$ at small $\xtwo$. 

\begin{figure}[htbp]
    \centering
    \includegraphics[scale=0.45]{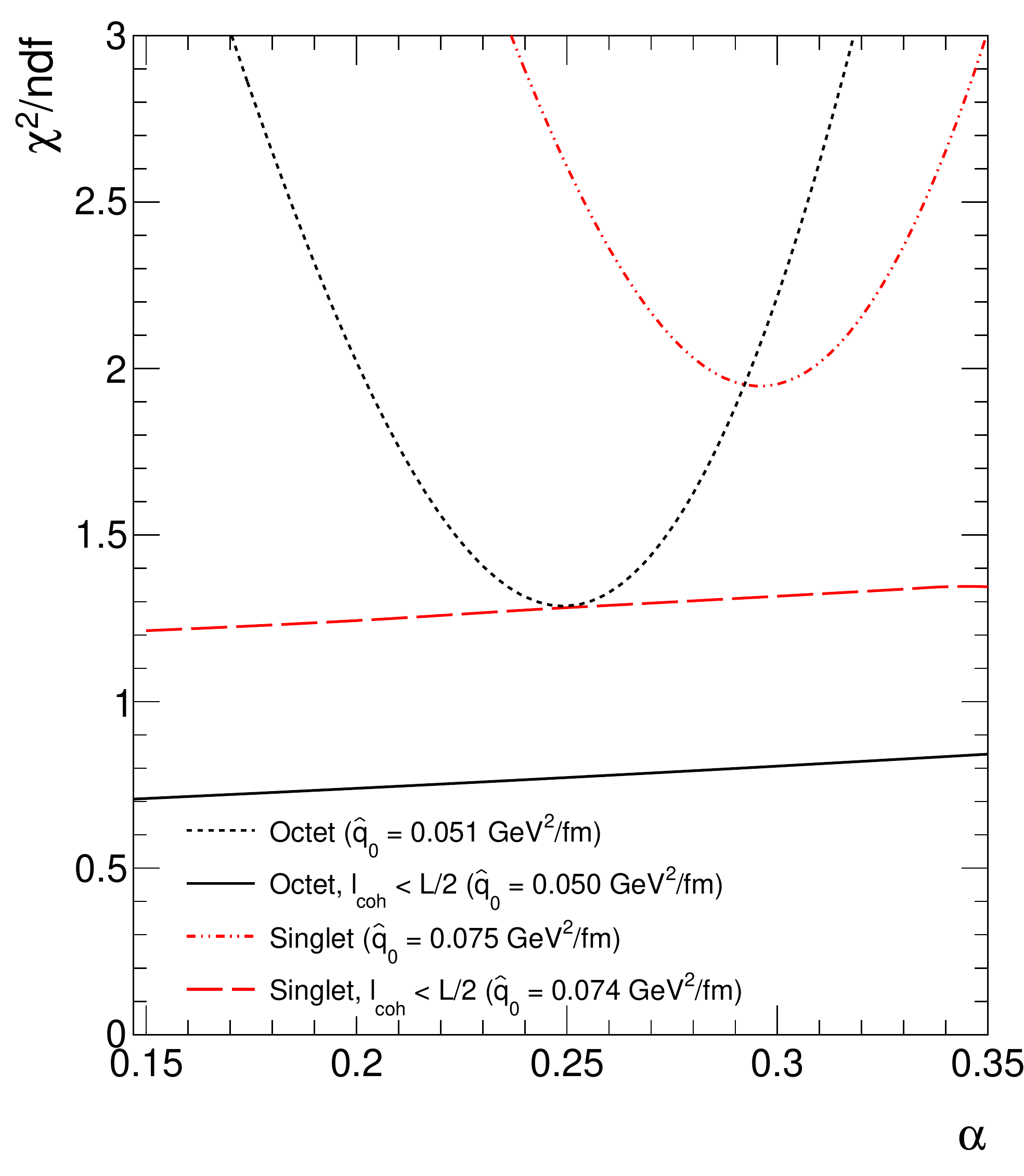}
    \caption{$\chindf$ profile from the fits to all and low $\lcoh<L/2$ data, assuming $Q\bar{Q}$ pairs to be produced in a color octet (respectively, dotted and solid) and color singlet (dash-dotted and dashed) state.}
    \label{fig:chi2profiles}
\end{figure}

Let us now discuss the values of the two parameters of the model obtained from the fit to the data. The best fit ($\chindf=1.3$) leads to the small-$\xtwo$ exponent of the transport coefficient $\alpha = 0.25 \pm 0.01$, as can be seen from the $\chindf$ profile shown as a dotted line in Fig.~\ref{fig:chi2profiles}. This value of $\alpha$ proves in very good agreement with the `geometrical scaling' fits of HERA data~\cite{GolecBiernat:1999qd}, which is quite remarkable given the different observables. It is also nicely compatible with the increase of the saturation scale deduced from elliptic flow measurements in heavy ion collisions from RHIC to LHC~\cite{Giacalone:2019kgg}. Recently a study based on the higher-twist framework, aiming at the extraction of the transport coefficient from a  global fit of semi-inclusive DIS and pA collisions data led to a slightly smaller exponent, $\qhat(\xtwo)\propto \xtwo^{-0.17}$~\cite{Ru:2019qvz}.

The absolute value of the transport coefficient obtained from the fit, $\qzero=0.051\pm0.002$~\gevsqfm, coincides with the perturbative estimate by BDMPS in Ref.~\cite{Baier:1996sk}. It is also consistent with estimates from fully coherent energy loss effects~\cite{Arleo:2012rs}.\footnote{The extracted transport coefficient assuming FCEL effects only leads to $\qzero=0.07$--$0.09$~\gevsqfm, that is slightly above the present estimate. Note however than when assuming fully coherent energy loss and nuclear parton densities effects, it was found that $\qzero=0.046$~\gevsqfm (with EPS09~\cite{Eskola:2009uj}) and $\qzero=0.064$~\gevsqfm (with DSSZ~\cite{deFlorian:2011fp}).} The quoted uncertainties on the two parameters follow from the $\chi^2$ analysis with the strict tolerance criterion of $\Delta \chi^2=+1$. However, it is clear that these uncertainties should be taken as lower estimates, as other sources of `systematic' uncertainties inherent in the model are likely to affect these results. One such assumption is the value of $x$ at which the transport coefficient is frozen, $x_{\A} = 1/(2 m_{_{\textnormal{N}}} L_{\A}$), see~\eqref{eq:qhatmodel}. Yet motivated on physical grounds by the uncertainty principle, this expression gives at most the magnitude for $x_{\A}$. Varying $x_{\A}$ by a factor of two with respect to that estimate leads respectively to $\qzero=0.055 \pm 0.002$~\gevsqfm and $\qzero=0.044 \pm 0.001$~\gevsqfm, still with a rather good agreement with data ($\chindf=1.9$ and $\chindf=1.4$, respectively), thus leading to an additional uncertainty on \qzero of about 10\%. It has also been checked that evaluating $\qhat(\xtwo)$ at different values of $\xtwo$, \eg $\xtwo=\left(M^2+\langle\pt^2\rangle\right)^{1/2}/\sqrt{s}\times e^{-y}$ at various $\langle\pt^2\rangle$, affects only marginally (by less than $5\%$) our results. The largest source of uncertainty on $\qzero$ is related to the assumption on the color state of the propagating $Q\bar{Q}$ state; it is discussed more specifically in section~\ref{sec:colorsinglet}.

It has been pointed out that the broadening may have a different expression at large coherence length for specific QCD processes, typically when the color charge of the incoming parton and that of the outgoing particle differ (\eg Drell-Yan or color singlet $Q\bar{Q}$ production) while \eqref{eq:broadeningpa} should hold at any coherence length in the case of color octet $Q\bar{Q}$ production, as assumed here. Nevertheless, in order to check the consistency of our approach at low and at high energy, a fit to a subset of data at small coherence length has been performed, taking as a criterion $\lcoh < L/2$ which excludes the LHC $\jpsi$ data and the RHIC $\jpsi$ measurements at mid and forward rapidity. The fit leads to $\qzero=0.050\pm0.002$~\gevsqfm ($\chindf=0.8$ with $\alpha=0.25$), a value which is nicely consistent with results from the full data sets, showing that the RHIC and LHC data do not alter the value extracted from low energy measurements. 

High energy data are nevertheless key in order to extract precisely $\alpha$, the small-$\xtwo$ exponent of the transport coefficient. Indeed, at low energy the transport coefficient \eqref{eq:qhatmodel} is independent of $\xtwo$ and solely depends on $\xiA \equiv 1/(2 m_{_{\textnormal{N}}} L)$.\footnote{Assuming $\lcoh < L/2$ actually leads to $\xtwo > 2\xiA$.} Since most of the nuclei involved in this analysis have similar atomic masses hence similar medium lengths ($A\simeq 200$ corresponding to $\xiA\simeq 10^{-2}$), the transport coefficient used in the low-energy fit is a constant independent of $\alpha$, $\qhat(x)\simeq \qzero$. These low-energy data alone are thus unable to constrain the parameter $\alpha$, as can be seen in Figure~\ref{fig:chi2profiles} (solid line) which shows a rather flat $\chindf$ profile as a function $\alpha$. The results of the fits, either to the full data set or to the small-$\lcoh$ subset of data, are summarized in Table~\ref{tab:fitresults}.
\begin{table}[htbp]
  \begin{center}
    \begin{tabular}{p{3cm}p{3cm}ccc}
      \hline 
      \hline 
       & Data set & $\qzero$ (\gevsqfm) & $\alpha$ & \chindf  \\
      \hline
    \multirow{2}{*}{Color octet}& All  &  $0.051 \pm 0.002$ & $0.25\pm0.01$ & $1.3$ \\
      & $\ell_c < L/2$  &  $0.050\pm0.002$ & $(0.25)$ & $0.8$ \\
      \hline
      \multirow{2}{*}{Color singlet} & All  & $0.075\pm0.003$ & $0.30\pm0.02$ & $2.0$\\
      & $\ell_c < L/2$  &  $0.074\pm0.003$ & $(0.30)$ & $1.3$ \\
      \hline 
      \hline
    \end{tabular}
     \caption{Results of the fits to all and selected ($\ell_c < L/2$) data sets, assuming color octet or color singlet $Q\bar{Q}$ production.}
    \label{tab:fitresults}
  \end{center}
\end{table}

\begin{figure}[htbp]
    \centering
    \includegraphics[scale=0.55]{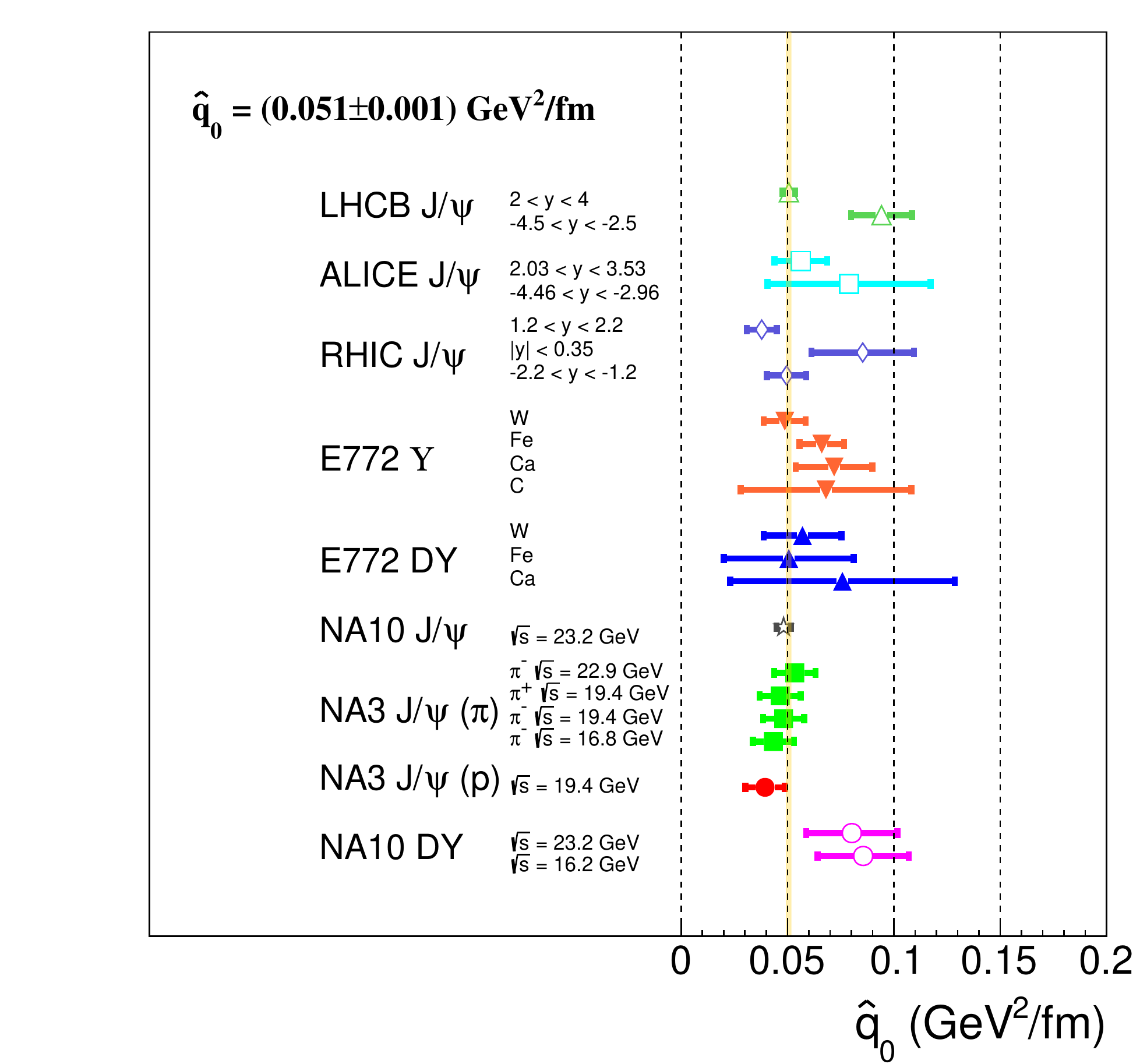}
    \caption{Extracted values of $\qzero$ from each measurement of $\dptsq$. Experiments are plotted in descending order of $\sqrt{s}$ energy, data points in ascending order of atomic number (E772) and rapidity (PHENIX, ALICE, LHCb).}
    \label{fig:q0}
\end{figure}
On top of the global fits, the values of $\qzero$ have been extracted from each data point using \eqref{eq:broadeningpa} and \eqref{eq:qhatmodel}. Within each experiment, the different values correspond either to different nuclei (E772), different collision energies (NA3, NA10), and to different rapidity bins (PHENIX, ALICE, LHCb), see Table~\ref{tab:datasets}.
Results plotted in Figure~\ref{fig:q0} show a remarkable consistency (as could be anticipated from Figure~\ref{fig:scalingCEM}), pointing to a common value for the transport coefficient $\qzero$. The weighted average of $\qzero$ is also consistent with the estimates above. It is interesting to note that the Drell-Yan measurements by NA10 lie slightly above the average. The largest tension is observed with the LHCb measurement at backward rapidity ($-4 < y < -2$), which might be partly attributed to FCEL or nPDF effects, as discussed in Appendix~\ref{app:fcelnpdf}.

\subsection{Relationship with the gluon distribution} \label{section:xGx}

As discussed in section~\ref{sec:qhat}, the transport coefficient is directly proportional to the gluon distribution of the nucleus, \eq{eq:qhatBDMPS}, which led to~\eq{eq:qhatmodel}, tested in the previous section. The large range of variation in the value of $x$, from $x \simeq 2\times10^{-2}$ at low collision energy to $x \simeq 2\times 10^{-5}$ at the LHC at forward rapidity,\footnote{Note, however, that most of the data points at low collision energies correspond to similar values of $x$, $x = x_\A$, see \eqref{eq:qhatmodel}.} makes it possible to check further the relationship between $\qhat$ and $xG(x)$ by comparing data to actual proton PDF sets. From \eqref{eq:qhatBDMPS}, the broadening reads
\be\label{eq:dptsq_xG}
\dptsq \propto {\cal C} \alpha_{s}(Q^2)\,\left[x G(x, Q^2)L_{\textnormal{A}} - x^\prime G(x^\prime, Q^2)L_{\textnormal{p}} \right] \equiv {\cal C}\,\alpha_{s}\,\Delta(xG\,L)\,,
\ee
where the proportionality constant is related to $\qzero$ which is now the only parameter, and $x^\prime\equiv\min(\xip, \xtwo)$. As mentioned in section~\ref{section:xGx}, the scale $Q^2$ appearing in \eqref{eq:dptsq_xG} is expected to be of the order of the broadening itself, $Q^2 = \lambda\, \dptsq$ with $\lambda=\cO{1}$. At such semi-hard scales (look at Figure~\ref{fig:scalingCEM}, in which most data points lie below $\dptsq \lesssim 1\,\gev^2$), parton densities are not defined and $Q^2$ should be frozen at the input scale, $Q^2 = \min(Q_0^2, \lambda \dptsq)$ defined in the global fit analysis (typically, $Q_0^2\sim1$--2~GeV$^2$). 
\begin{figure}[tbp]
\centering
    \includegraphics[scale=0.55]{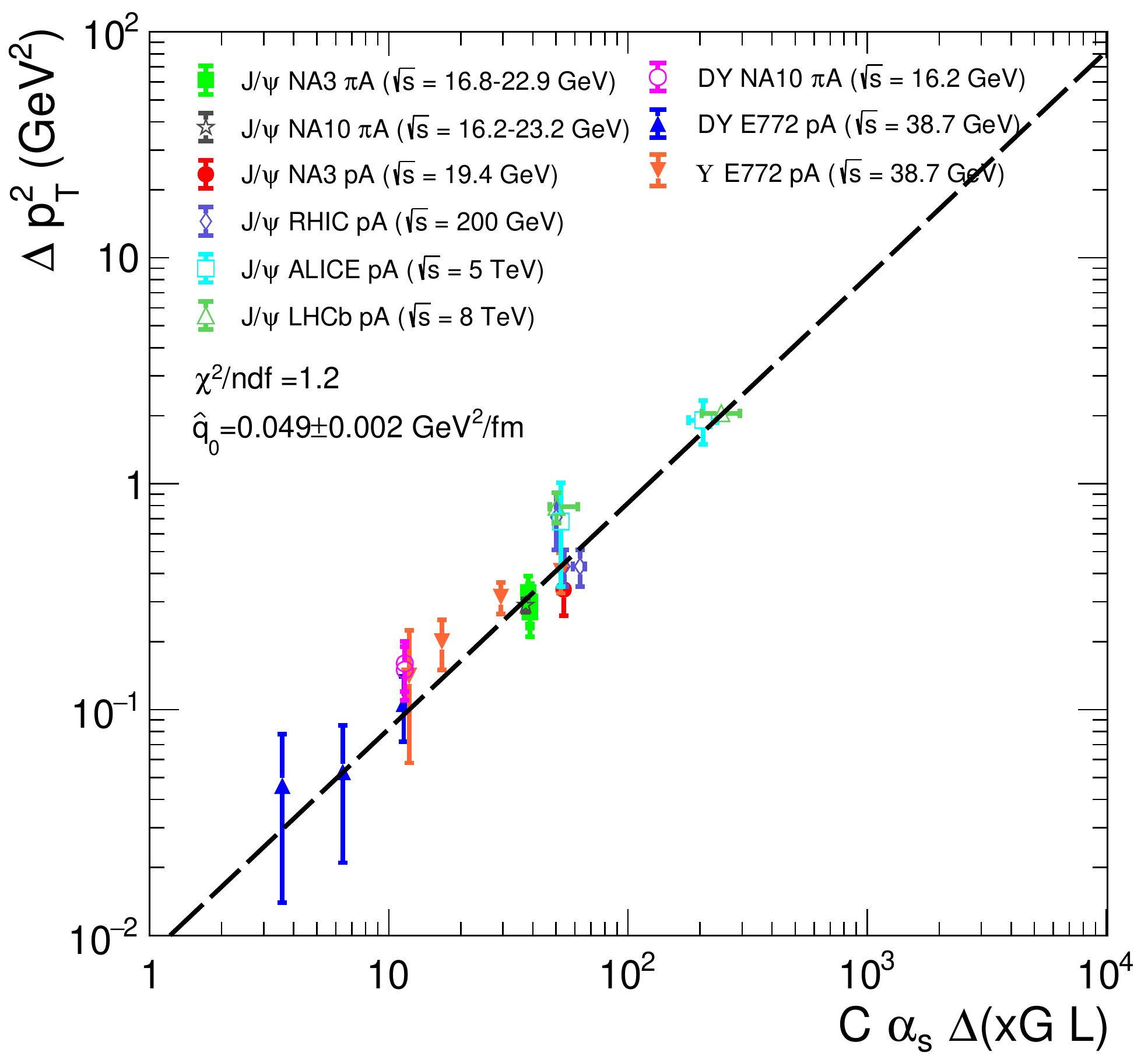}
    \caption{Scaling of the nuclear $\pt$-broadening in DY and quarkonium production using the CT14 LO gluon distribution.}
    \label{fig:xGx}
\end{figure}

The fit has been performed using the leading-order\footnote{The choice of a LO PDF is consistent with the leading-order relationship between the broadening and the transport coefficient, \eq{eq:qhatBDMPS}.} CT14 PDF set~\cite{Dulat:2015mca}. A bit surprisingly, the choice $\lambda=1$ leads to a rather poor agreement with data. Due to the quasi-absence of QCD evolution from $Q_0^2$ to $\dptsq$, the gluon distribution does not rise sufficiently at small $x$ to account for the measurements. Better agreement is nevertheless found for larger values of $\lambda$, \eg $\lambda \simeq 2.5$ ($\chindf=1.2$) as can be seen in Figure~\ref{fig:xGx} which compares data and the assumption~\eqref{eq:dptsq_xG}. Moreover the obtained value for the transport coefficient at $x=10^{-2}$, $\qzero=0.049\pm0.002$~\gevsqfm, is compatible with our previous estimate based on a simpler model for the gluon distribution.

\subsection{Color singlet production}\label{sec:colorsinglet}

The present lack of understanding regarding the  production process of quarkonium states makes it difficult to predict the nuclear broadening of $\jpsi$ and $\Upsilon$ states. It has been assumed in this analysis that the $Q\bar{Q}$ pair neutralizes its color on a long timescale, as in the CEM or NRQCD, and for which \eq{eq:broadeningpa} appears appropriate at both low and high energies. In this section, we shall assume that $Q\bar{Q}$ states turn color singlet on a short timescale, in the spirit of the color singlet model (CSM). In the CSM, the leading order process for quarkonium production is gluon fusion, thus leading to the color factor ${\cal C}=N_c/2$ in both pA and $\pi$A collisions\footnote{In CEM and NRQCD, gluon fusion and quark-antiquark annihilation processes occur at the same order in the perturbative expansion, but parton distributions favor the former in \pA collisions and the latter in $\pi$A collisions, see discussion in section~\ref{sec:processdependence}.} (see Table~\ref{tab:color}). 

As \eq{eq:broadeningpa} may not hold at large coherence length in the case of color singlet $Q\bar{Q}$ production, we first compare the model expectations to the DY and quarkonium data at small coherence length, $\lcoh < L/2$. The fit of this subset of data leads to a transport coefficient $\qzero=0.074\pm0.003$~\gevsqfm ($\chindf=1.3$ at $\alpha=0.30$). Without much surprise, the transport coefficient extracted assuming color singlet $Q\bar{Q}$ production proves somewhat larger than assuming color octet $Q\bar{Q}$ pairs, in order to compensate for the smaller color factor, see Table~\ref{tab:color}. As in the case of color octet $Q\bar{Q}$ production, these low energy data alone do not allow for the determination of the small-$\xtwo$ exponent $\alpha$; see the flattish $\chindf$ profile in Fig.~\ref{fig:chi2profiles} (dashed line). For completeness, a fit of the full data set has been performed under the color singlet assumption. The fits leads to $\qzero=0.075\pm0.003$~\gevsqfm (however with a rather large $\chindf=2.0$), which is also consistent with the value obtained from low energy data. The use of the whole data set now allow for a precise extraction of $\alpha$, $\alpha=0.30\pm0.02$ (Figure~\ref{fig:chi2profiles}, dash-dotted line). Interestingly, this estimate is comparable in magnitude with our earlier result assuming color octet $Q\bar{Q}$ states, and also compatible with small-$x$ deep inelastic scattering data~\cite{GolecBiernat:1999qd}. This result should however be taken with care as \eq{eq:broadeningpa} may not be valid at high energy in the case of color singlet production.

\section{Conclusion}\label{sec:conclusion}

The transverse momentum nuclear broadening of Drell-Yan lepton pair and quarkonium production has been investigated systematically, for different nuclear targets and different systems (pA and \piA collisions), from SPS to LHC energy. Within a model that includes respectively the dependence on the medium length, the process-dependent color factors, and the $x$ dependence of the transport coefficient (or equivalently that of the saturation scale), a simple scaling is expected and is observed in the data.

This allows for the determination of the transport coefficient of cold nuclear matter at $x=10^{-2}$, $\qzero=0.051$~\gevsqfm, with a systematic uncertainty estimated to be 10\%.  Moreover, the best fit to data points to the $x$-dependence, $\qhat(x) \sim x^{-0.25}$. This would correspond to a saturation scale in a large nucleus (either Au or Pb) of $Q_s=0.7$~GeV at RHIC at mid-rapidity and $Q_s=1.1$~GeV at LHC at mid-rapidity (using here $\xtwo\simeq M_{\jpsi}/\sqrts$).
The relation expected in perturbative QCD between the transport coefficient and the gluon distribution in a proton is explored further using CT14 LO gluon distribution. Good agreement is reported provided the hard scale entering $xG(x, Q^2)$ and $\alphas(Q^2)$ is $Q^2 \simeq 2.5\,\dptsq$, that is slightly above the BDMPS prescription $Q^2 = \dptsq$~\cite{Baier:1996sk}.
The analysis has also been carried out assuming that $Q\bar{Q}$ pairs turn color singlet on a short timescale. In this case, the full dataset would favor a naturally larger transport coefficient, $\qzero=0.075$~\gevsqfm, albeit with a similar small-$x$ exponent, $\alpha=0.3$.

The present picture could be further tested with additional measurements. At the LHC, Drell-Yan data at backward/forward rapidity by LHCb and at midrapidity by ATLAS and CMS would be extremely valuable. These data would shed light on the DY broadening at small values of $\xtwo$, that is when the coherence length for the hard process is significantly larger than the medium length (the RHIC forward DY data may also be important in this respect~\cite{Leung:2018tql}). In addition, at lower energy, the simultaneous measurement of DY and \jpsi nuclear broadening in \piA collisions at forward rapidity by the COMPASS experiment~\cite{Aghasyan:2017jop}, or in pA collisions at backward rapidity by the LHCb-SMOG experiment~\cite{Bursche:2649878,Aaij:2018ogq}, should also bring essential information on the value of the transport coefficient as well as on quarkonium formation dynamics.

\appendix
\section{FCEL and nPDF effects on \texorpdfstring{$\bm{\Delta p_\perp^2}$}{broadening}}
\label{app:fcelnpdf}

The effects of FCEL and nPDF on the shape of $\jpsi$ \pt-spectra in \pA collisions, hence on the nuclear broadening $\dptsq$, are investigated. Neglecting multiple scattering and focusing on the sole FCEL/nPDF effects, the cross section in \pA collisions can be modelled as (see section~\ref{sec:fcelnpdf}),
\begin{align}\label{eq:fcelnpdf}
\frac{1}{A}\,\frac{\dd\sigma_\pA^{\textnormal{FCEL}}(\pt, y)}{\dd \pt \dd y} 
&= \rpa^{\textnormal{FCEL}}\left(\pt, y\right) \times \frac{\dd\sigma_\pp(\pt, y)}{\dd \pt \dd y}\,, \\
\frac{1}{A}\,\frac{\dd\sigma_\pA^{\textnormal{nPDF}}(\pt, y)}{\dd \pt \dd y}
&= R_g^\A\left(\xtwo=\frac{M_\perp+\pt}{\sqrt{s}}\,e^{-y}, Q=M_\perp+\pt\right) \times \frac{\dd\sigma_\pp(\pt, y)}{\dd \pt \dd y}\,. \nonumber
\end{align}
The FCEL quarkonium nuclear production ratio $\rpa^{\textnormal{FCEL}}$ is computed from Ref.~\cite{Arleo:2012rs}, while the gluon nPDF ratio $R_g^{\A}$ is given by EPPS16~\cite{Eskola:2016oht}. The choice made here for $\xtwo$ and $Q$ that enter $R_g^\A$ interpolates between the $2\to1$ and $2\to2$ kinematics expected at low and high $\pt$.
The double differential $\jpsi$ production cross section in \pp collisions is parametrized as
\begin{equation}\label{eq:ppxs}
\frac{\dd\sigma_{\pp}(\pt, y)}{\dd{y}\,\dd\pt}\propto \pt\times\left(\frac{p_{0}^2}{p_{0}^2+\pt^2}\right)^{m} \times\left(1-\frac{2 M_{\perp}}{\sqrt{s}} \cosh y\right)^{n}\,,
\end{equation}
where the values of the parameters $\mathcal{N}$, $n$, $m$ and $p_0$ were extracted from $\sqrts = 200$~GeV and $\sqrts = 7$~TeV data in Ref.~\cite{Arleo:2013zua}.
Using the spectrum \eqref{eq:ppxs} in \eqref{eq:fcelnpdf} allows for computing $\dptsqfcel$ and $\dptsqnpdf$. Calculations are performed at RHIC and at LHC ($\sqrt{s}=8.16$~TeV), in the rapidity acceptance of the PHENIX experiment ($1.2 < |y| < 2.2$ and $|y| < 0.35$) and the LHCb experiment ($-4.5 < y < -2.5$ and $2 < y < 4$), respectively.\footnote{In the case of FCEL calculations,  $\rpa^{\textnormal{FCEL}}$ is computed in the median of each rapidity bin.}

Results are summarized in Table~\ref{tab:fcelnpdf}. At RHIC, the values of $\dptsq$ due to FCEL are small, $\dptsqfcel=0.1$~GeV$^2$ in all rapidity bins. At the LHC, FCEL alone leads to transverse momentum broadening ranging from $\dptsq=0.2$~GeV$^2$ at backward rapidity ($y=-3.5$) to $\dptsq=0.5$--$0.6$~GeV$^2$ at forward rapidity ($y=3$), consistent with the fact that FCEL effects are more pronounced in the proton fragmentation region. The quoted uncertainties arise from the variation of the transport coefficient from $\qzero=0.050$ to $\qzero=0.075$~\gevsqfm, consistently with the results obtained in this analysis.

Let us now comment on the nPDF effects. At RHIC, the nPDF contribution to $\dptsq$ at backward rapidity is slightly negative as the region between anti-shadowing and the EMC effect softens \pt-spectra in \pA collisions with respect to \pp collisions (see discussion in section~\ref{sec:fcelnpdf}). The quoted uncertainty is determined after the variation of all EPPS16 nuclear member sets. At mid-rapidity and at forward rapidity, $\dptsqnpdf$ is positive in the two rapidity bins ($0.1 < \dptsqnpdf < 0.4$~GeV$^2$). At LHC, the values of $\dptsq$ range from $\dptsq=0.1$ to $\dptsq=0.6$~GeV$^2$ at backward rapidity and from $\dptsq=0.1$ to $\dptsq=0.7$~GeV$^2$ at forward rapidity, with a significant correlation observed for each EPPS16 member set between the two rapidity intervals (correlation coefficient of $0.7$).
\begin{table}[ht!]
  \begin{center}
    \begin{tabular}{p{2.5cm}cccc}
      \hline 
      \hline 
      Experiment & System & $y$ range & $\dptsqfcel$ (GeV$^2$) & $\dptsqnpdf$ (GeV$^2$)\\[0.08cm]
        \hline 
       PHENIX & dAu & $-2.2<y<-1.2$ & 0.1 & [-0.2 ; 0]  \\
     & dAu & $|y|<0.35$ & 0.1 & [0.1 ; 0.4]  \\
     & dAu & $1.2<y<2.2$ & 0.1 & [0.1 ; 0.4]  \\
        \hline 
       LHCb & pPb & $-4.5<y<-2.5$ & 0.2 & [0.1 ; 0.6] \\
     & pPb & $2<y<4$ & [0.5 ; 0.6] & [0.1 ; 0.7]  \\
      \hline 
      \hline
    \end{tabular}
     \caption{Calculation of $\dptsqfcel$ and $\dptsqnpdf$ at RHIC and LHC.}
    \label{tab:fcelnpdf}
  \end{center}
\end{table}

Although they might a play a role on the values of $\dptsq$, this study reveals that both FCEL and nPDF effects appear to be small (yet with a large uncertainty in the latter case) compared to the values observed in data, suggesting that multiple scattering is the leading effect, as assumed throughout the present analysis.

\section{Extraction of \texorpdfstring{$\bm{\Delta p_\perp^2}$}{broadening} at RHIC and LHC}
\label{app:broadening}

In practice, the average transverse momentum squared defined as
\begin{equation}
\ptsq = {\int_{0}^{\infty} \dd\pt\, \pt^2\, \frac{\dd\sigma}{\dd\pt}}\, \Bigg/ \, {\int_{0}^{\infty} \dd\pt\, \frac{\dd\sigma}{\dd\pt}}\,,
\label{eq:ptsqint}
\end{equation}
cannot be extracted from data. Instead, $\ptsq$ is often determined from the sum over the experimental bins (with lower and upper edges, $\pt^{i-1}$ and $\pt^{i}$),
\begin{equation}
{\ptsq}_{\rm bins} = \sum_{i=1}^{N_{\textrm{bins}}} {(\pt^{i} - \pt^{i-1})\times \left(\hat{p}_{\perp}^i\right)^2\, \frac{\dd\sigma^i}{\dd\pt}}\; \Big/\; \sum_{i=1}^{N_{\textrm{bins}}} {(\pt^{i} - \pt^{i-1})\times \frac{\dd\sigma^i}{\dd\pt}}\,,
\label{eq:ptsqbins}
\end{equation}
taken as an approximation of \eq{eq:ptsqint}. This latter expression has however two drawbacks. When the bins are too wide, as is often the case at large \pt, the typical $\hat{p}_\perp^i$ value at which the transverse momentum should be evaluated in the bin $[\pt^{i-1}, \pt^{i}]$, for instance the average or the median in that bin, may significantly affect the value of ${\ptsq}_{\rm bins}$. In addition, \eq{eq:ptsqbins} should be used only when the highest \pt value reached in the experiment, $\pt^{\max}=\pt^{N_{\textrm{bins}}}$, is large enough so that ${\ptsq}_{\rm bins}$ is independent of this upper cutoff, within the experimental uncertainty.

Whenever available, we use in the present article the values of $\dptsq$ published in the experimental analyses. This is the case of all fixed-target experiments. The specific case of PHENIX data is specifically discussed hereafter. At the LHC the values of $\dptsq$ in minimum bias pPb collisions have not been published, neither by ALICE nor by LHCb. The extraction of $\dptsq$ from these data is discussed below.\\

\noindent {\bf PHENIX --} The PHENIX experiment published $\ptsq$ values for \jpsi production in pp and dAu collisions at $\sqrts = 200$~GeV in different rapidity bins~\cite{Adare:2012qf}. Using the absolute cross sections measured in pp and dAu collisions~\cite{Adare:2012qf,Adare:2011vq}, we have checked that the broadening values quoted in~\cite{Adare:2012qf} can be recovered when using the median transverse momentum, $\hat{p}_\perp^i=(\pt^{i-1}+\pt^{i})/2$, in \eq{eq:ptsqbins}. However, because of the wide experimental bins at large $\pt$, the values of ${\ptsq}_{\rm bins}$ in pp and dAu collisions, hence the nuclear broadening $\dptsq$, is significantly affected when replacing the median by the \emph{average} transverse momentum in each bin\footnote{This value is not given in~\cite{Adare:2012qf} but was estimated using the fit~\eqref{eq:kaplan} in each bin.} in \eq{eq:ptsqbins}. In order to circumvent the drawbacks of using \eq{eq:ptsqbins}, data have been fitted using the so-called Kaplan distribution,
\begin{equation}
\frac{\dd\sigma_{\textnormal{fit}}(\pt)}{\pt\dd\pt}=\mathcal{N}\left(\frac{p_{0}^2}{p_{0}^2+\pt^2}\right)^{m}
\label{eq:kaplan}
\end{equation}
named after Ref.~\cite{Kaplan:1977kr}. Once the parameters $m$ and $p_0$ are known (the parameter ${\cal N}$ is irrelevant when computing $\ptsq$), the value of $\dptsq$ can be determined using the proper definition \eq{eq:ptsqint}.\\

\begin{table}[tbp]
  \begin{center}
    \begin{tabular}{p{3.cm}cc}
      \hline 
      \hline 
      Experiment & $y$ range & $\dptsq$ (GeV$^2$) \\
     \hline 
     PHENIX &  $-2.2<y<-1.2$ & 0.43 $\pm$ 0.08\\
     &  $|y|<0.35$ & 0.71 $\pm$ 0.20 \\
     & $1.2<y<2.2$ & 0.43 $\pm$ 0.08 \\
     \hline 
     LHCb & $-4.5<y<-2.5$ & 0.79 $\pm$ 0.12 \\
     &  $2<y<4$ & 2.05 $\pm$ 0.12\\
      \hline 
     ALICE &  $-4.46<y<-2.96$ & 0.68 $\pm$ 0.33 \\
     &  $2.03<y<3.53$ & 1.91 $\pm$ 0.42 \\
      \hline 
      \hline
    \end{tabular}
     \caption{Determination of $\dptsq$ from PHENIX, LHCb and ALICE measurements.}
    \label{table:LHC_value}
  \end{center}
\end{table}

\noindent {\bf LHCb --} The same procedure, namely using \eqref{eq:kaplan} in \eqref{eq:ptsqbins}, is applied to \jpsi production cross sections measured by LHCb in pp and pPb collisions at $\sqrts = 8.16$~TeV \cite{Aaij:2017cqq}.\\

\noindent {\bf ALICE --} The ALICE experiment has released the values of $\dptsq$ for \jpsi production in pPb (and pp) collisions at $\sqrts = 5.02$~TeV~\cite{Adam:2015jsa} in different centrality bins. Here the weighted average over the $N$ centrality bins has been carried out,
\begin{equation}
{\ptsq}_{\textnormal{Pb}} = \sum_{\cal C} {\ptsq}_{\cal C}\, \sigma_{\cal C}\; \Big/\;  \sum_{\cal C} \sigma_{\cal C}
\end{equation}
where $\sigma_{\cal C}$ are the measured \pt-integrated \jpsi production cross section in each centrality bin.

The values of $\dptsq$ extracted from PHENIX, LHCb and ALICE are given in Table~\ref{table:LHC_value}.

\acknowledgments
We would like to thank St\'ephane Peign\'e and St\'ephane Platchkov for useful comments and discussions. This work is funded by ``Agence Nationale de la Recherche'' under grant COLDLOSS (ANR-18-CE31-0024-02).

\providecommand{\href}[2]{#2}\begingroup\raggedright\endgroup

\end{document}